\documentclass[aps, pra, twocolumn, floatfix, amsmath, amssymb, nofootinbib, superscriptaddress, oneside]{revtex4}

\usepackage[german,english]{babel}
\usepackage[utf8]{inputenc}
\usepackage{amsmath,amssymb}
\usepackage{amsthm}
\usepackage{paralist}
\usepackage{algorithmic}
\usepackage{graphicx}
\usepackage{mathtools}
\usepackage{url}
\usepackage[babel,german=quotes]{csquotes}
\usepackage{simplewick}
\usepackage{slashed}
\usepackage{hyperref}
\usepackage{color}
\usepackage{braket}
\usepackage{gensymb}
\usepackage{siunitx}
\usepackage{bbm}
\usepackage{mhchem}
\usepackage{physics}
\usepackage{tabularx}
\usepackage[caption=false, font=small, singlelinecheck=false]{subfig}
\usepackage[section]{placeins}
\theoremstyle{definition}  
\theoremstyle{definition}  
\theoremstyle{remark}      
\theoremstyle{remark}      
\theoremstyle{remark}

\begin{document}

\title{Vacuum Polarization Effects in the Hyperfine Splitting of Hydrogen Like Ions}

\author{J. Heiland Hoyo}
\affiliation{Max-Planck-Institut f\"{u}r Kernphysik, Heidelberg, Germany}
\affiliation{Ruprecht Karl University of Heidelberg, Department of Physics and Astronomy, Heidelberg, Germany}
\author{B. Sikora}
\affiliation{Max-Planck-Institut f\"{u}r Kernphysik, Heidelberg, Germany}
%wer noch?

\begin{abstract}
    The hyperfine structure of bound electrons in hydrogen-like ions is considered with corrections to the energy levels due to vacuum polarization (VP). Corrections to the wave function as well as the magnetic potential are determined for both leptonic and hadronic VP. Hadronic VP is treated with a semi-empirical approach. Uncertainties due to the nuclear charge distribution are given and discussed. Point-like, spherical and Fermi distributed nuclear models are considered and the differences of the results are discussed.
\end{abstract}

\date{\today}

\maketitle

\section{Introduction}

Few-electron atomic systems are an ideal tool to test quantum electrodynamics (QED) in the presence of strong electric background fields~\cite{CODATA,PlunienSoff,Shabaev2002}. Stringent QED tests have recently been performed by comparing experimental and theoretical values for the bound electron's $g$-factor~\cite{Sturm11,Schneider2022,Morgner2023}. The established value of the electron mass stems from the bound-electron $g$-factor of the hydrogenlike $^{12}$C$^{5+}$ ion~\cite{Sturm14,CODATA}.

If the atomic nucleus has a non-zero spin, the ground-state energy level of the bound electron splits into two sublevels due to the magnetic interaction between the electron's and the nuclear spin. This is called the hyperfine splitting (HFS). The leading (zero-loop) contribution to the HFS, including relativistic effects, is known analytically (e.g.~\cite{Beier,Shabaev1994}). Corrections due to the finite size nuclear charge and magnetization distributions (the latter being referred to as the Bohr-Weisskopf correction) are sizeable even for the lightest elements~\cite{Volotka2005}. The accuracy of the theoretical value of the HFS is typically limited by the Bohr-Weisskopf correction~\cite{Shabaev1994}. This has lead to the extraction of nuclear properties, specifically the Zemach radii, from HFS measurements, rather than directly comparing experimental and theoretical HFS values. Zemach radii have been extracted for the proton~\cite{Volotka2005}, as well as several Li isotopes Li~\cite{Pachucki2023Li} and $^9$Be~\cite{Pachucki2014}. The Zemach radius of the $^3$He nucleus obtained from HFS~\cite{Schneider2022,Patkos2023} was found to be in tension with the Zemach radius obtained from elastic electron scattering data~\cite{Sick2014}. 

Apart from determining nuclear parameters, a weighted difference of hyperfine splitting values of the H-like and Li-like charge states allows the suppression of nuclear effects and therefore tests of atomic theory. For the weighted difference of hyperfine splittings in $^{209}$Bi ions, a 7$\sigma$ discrepancy between theoretical and experimental values was initially found~\cite{Ullmann2017}. This discrepancy was later resolved by a new measurement of the magnetic moment of the $^{209}$Bi nucleus~\cite{Skripnikov2018}. More recently, a good agreement was found between theoretical and experimental values for a weighted difference of hyperfine splittings in $^9$Be~\cite{Dickopf2024}.

Apart from nuclear corrections, Feynman diagrams with closed loops (radiative corrections) need to be considered. One-loop self-energy (SE) corrections have been calculated for both point-like~\cite{Yerokhin2010oneloop} and extended nuclei~\cite{Beier}. Also, one-loop vacuum polarization (VP) corrections have been calculated in the literature~\cite{Beier,KarshenboimLepVP}. In these references, only the case of electronic VP (i.e. the case of virtual electron-positron pairs) is discussed. In this work, we will systematically study the cases of virtual muon-antimuon pairs as well as hadronic VP. These effects are found to be strongly suppressed compared to electronic VP due to the much higher virtual particle mass. However, a precise determination of nuclear properties from HFS measurements requires a precise knowledge of all possible standard model contributions, including the strongly suppressed heavy-particle VP effects.

In this paper, we systematically calculate both the muonic and hadronic VP corrections to the HFS. The hadronic VP function is calculated using the same semi-empirical approach that was previously used for the calculation of hadronic VP corrections to the binding energy~\cite{Friar,Eugen,karsh2021} and to the bound-electron $g$-factor~\cite{Dizer2023}. We also calculate the electronic VP correction to the HFS and compare with literature values, as a consistency test of our analytic and numerical methods.

This paper is organized as follows. In the following section, we introduce the different models of the nuclear charge distribution used in this work. In section III, we discuss the two cases of leptonic (i.e. electronic and muonic) and hadronic VP and its impact on the interaction (Coulomb) potential between electron and nucleus. In section IV, we introduce the formulas relevant for hyperfine splitting and discuss the parametrization we use for our results. Analytic and numerical calculation methods used for both the electric and magnetic loop VP corrections are discussed in section V. We derive approximation formulas, valid for low nuclear charge numbers $Z$, for both leptonic and hadronic VP, in section VI. Finally, our results are discussed in section VII.

\section{Finite Sized Nuclear Models}
For point-like nuclei, the wave functions resulting from solving the Dirac equation
\begin{equation}
    (\vb*{\alpha}\cdot\vb*{p}+m_e\beta+V(r))\psi(\vb*{r})=E\psi(\vb*{r}),
    \label{eq:de}
\end{equation}
are well known for a Coulomb potential~\cite{Greiner}. They can be written as
\begin{equation}
    \psi(r,\Theta,\psi)= 	\begin{pmatrix}\frac{1}{r} G_{n\kappa}(r) \Omega_{\kappa m}(\Theta,\phi) \\ \frac{i}{r} F_{n\kappa}(r) \Omega_{-\kappa m}(\Theta,\phi)\end{pmatrix}.
    \label{eq:wf}
\end{equation}
with a separation into radial components, $F_{n\kappa}(r)$ and $G_{n\kappa}(r)$ and angular components, also called the spherical spinors $\Omega_{\kappa m}$ and $\Omega_{-\kappa m}$ i.e.~\cite{Greiner}. The indices indicate the principal quantum number $n$, the relativistic angular momentum quantum number $\kappa$ and the magnetic quantum number $m$.

Assuming the nucleus to be point-like is however a great simplification that can not be taken for granted. There are different ways to model a more realistic finite-sized nucleus. Using the root mean square radius $r_\text{rms}$, we can achieve a simple and relatively accurate model by assuming the charge distribution $\rho(r)$ of a homogeneously charged sphere with effective radius $R=\sqrt{5/3\langle r_{\text{rms}}^2\rangle}$,
\begin{align}
    \rho(r)_{\text{sphere}}&=\frac{3Ze}{4\pi R^3}\Theta(R-r),
    \label{eq:sphere}
\end{align}
as well as the more realistic Fermi distribution with the two-parameter form, given by
\begin{align}
    \rho(r)_{\text{Fermi}}&=Ze\frac{N}{1+e^{(r-c)/a}}.
    \label{eq:fermi}
\end{align}
The parameters $c$ and $a$ denote the half-density radius and skin thickness, respectively. The latter is related to $t=4\ln{3a}$, which is the radial distance over which the charge density falls from 90\% to 10\% of its value at $r=0$. For most nuclei, the value $t\approx2.3$ fm is a good approximation~\cite{Beier}. The half-density radius can be calculated to a good approximation as
\begin{align}
    c^2=\frac{5}{3}r_{\text{rms}}^2-\frac{7}{3}a^2\pi^2.
    \label{eq:cFermiapp}
\end{align}
\begin{figure}[t]
    \centering
    \includegraphics[width=0.49\textwidth]{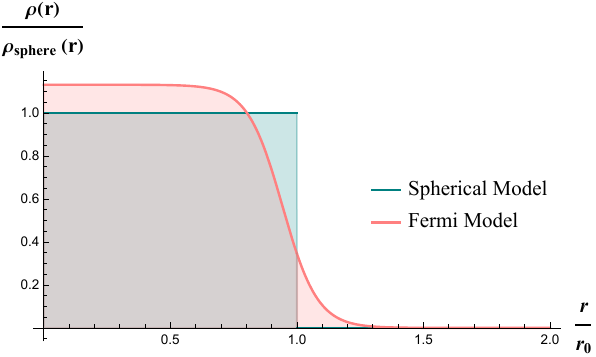}
    \caption{Plot of the charge densities for both models in case of $Z=92$}
\end{figure}
\noindent The exact value of $c$ as well as the normalization constant $N$ can be determined numerically.\\
The potential in case of a spherical model is given by 
\begin{equation}
    V(r)_{\text{sphere}} = 
    \begin{dcases*}
    -\frac{Z\alpha}{2R}\left(3-\frac{r^2}{R^2}\right) & $r \le R$ \\
    -\frac{Z\alpha}{2R} & $r > R$.
    \end{dcases*}
    \label{eq:potnew}
\end{equation}
For the Fermi model the potential is given in Ref.~\cite{FermiPot} by expanding the expression
\begin{align}
    -rV(r)=4\pi\left(\int_0^rds\ s^2\rho_{\text{Fermi}}(s)+r\int_r^{\infty}ds\ s\rho_{\text{Fermi}}(s)\right).
\end{align}

\section{Vacuum Polarization}
Following Ref.~\cite{Greiner2}, the modified nuclear static potential, also called the Uehling potential, for point-like nuclei can be written as 
\begin{align}
    \delta V_{\text{Ueh}}(r)=-\frac{2e}{\pi}\int_0^{\infty}dq\ j_0(qr)\text{Re}[\Pi^R(-q^2)]
    \label{eq:uehlAllg}
\end{align}
with the regularized polarization function $\Pi^R(-q^2)$ and the current $j_0(qr)$. It leads to a shift of atomic energy levels, which, at first order, is given by
\begin{align}
\begin{split}
    \Delta E_{n\kappa m}&=\langle n\kappa m | \delta V_{\text{Ueh}} | n\kappa m\rangle\\\
    &=\int_0^{\infty}dr \ \delta V_{\text{Ueh}}(r)[G_{n\kappa}^2(r)+F_{n\kappa}^2(r)]
\end{split}
\end{align}

Similarly to Ref.~\cite{Eugen}, the Uehling potential for an extended nucleus can be calculated by convoluting the point-like Uehling potential with the charge distributions in Eq.~(\ref{eq:sphere}) and (\ref{eq:fermi})
\begin{align}
        \delta V_{\text{fns}}(r)&=\frac{1}{Z\alpha}\int d^3x\ \rho(\vb*{x})\delta V_{\text{Ueh}}(\vb*{r}-\vb*{x}).
        \label{eq:ExtUeh}
\end{align}
\subsection{Leptonic Vacuum Polarization}
Leptonic VP is already well established in literature. Following the example established in Ref.~\cite{Peskin}, we can express the Uehling potential in Eq.~(\ref{eq:uehlAllg}) by expanding the integration contour into the imaginary part. This results in
\begin{align}
    \delta V_{\text{Ueh}}(r)=-\frac{2\alpha}{\pi r}\int_{2m}^{\infty}dq\frac{e^{-qr}}{q}\text{Im}[\Pi^R(-q^2\pm i\varepsilon)]
\end{align}
with the imaginary part of the leptonic vacuum polarization given by
\begin{align}
    \text{Im}[\Pi^R(-q^2)]=\frac{\alpha}{3}\sqrt{1-\frac{4m_{\text{lep}}^2}{q^2}}\left(1+\frac{2m_{\text{lep}}^2}{q^2}\right).
    \label{eq:PiLep}
\end{align}
\subsection{Hadronic Vacuum Polarization}
In case of hadronic vacuum polarization, a perturbative quantum chromodynamic approach fails due to the strong interaction having to be taken into account~\cite{Fred}. Another possibility to calculate it nevertheless is a semi-empirical approach using experimental data from $e^-e^+\rightarrow \text{hadrons}$ collisions. Following the approach in Ref.~\cite{Friar,Eugen} we express the real part of the polarization function with it's imaginary part and link that to the measurable total cross section $\sigma_{e^-e^+\rightarrow\text{hadrons}}$. In Ref.~\cite{Burkhardt} data from multiple experiments at different energy regions of center-of-mass collisions was used to construct an approximate parametrization of the polarization function
\begin{align}
    \text{Re}[\Pi^R_{\text{had}}(q^2)]=A_i+B_i\ln(1+C_i|q^2|)
\end{align}
with the constants $A_i,B_i,C_i$ defined for different regions of $q^2$. For our evaluation, we will be using an updated version for these constants, given in Tab.~\ref{tab:hadVPT}.
\begin{table}[h!]
    \centering
    \caption{Values for the parametrization of the hadronic vacuum polarization function with the mass of the Z boson $m_Z$ \cite{Burkhardt2}.}
    \begin{tabular*}{\linewidth}{@{\extracolsep{\fill}} ccccc }
    \hline
    \hline
        Region & Range (GeV) & $A_i$ & $B_i$ & $C_i$ ($\text{GeV}^{-2}$) \\
        \hline
        $k_0-k_1$ & 0.0-0.7 & 0.0 & 0.0023092 & 3.9925370\\
        $k_1-k_2$ & 0.7-2.0 & 0.0 & 0.0023333 & 4.2191779\\
        $k_2-k_3$ & 2.0-4.0 & 0.0 & 0.0024402 & 3.2496684\\
        $k_3-k_4$ & 4.0-10.0 & 0.0 & 0.0027340 & 2.0995092\\
        $k_4-k_5$ & 10.0-$m_Z$ & 0.0010485 & 0.0029431 & 1.0\\
        $k_5-k_6$ & $m_Z$-$10^4$ & 0.0012234 & 0.0029237 & 1.0\\
        $k_6-k_7$ & $10^4$-$10^5$ & 0.0016894 & 0.0028984 & 1.0\\
    \hline
    \hline
    \end{tabular*}
    \label{tab:hadVPT}
\end{table}\\
The resulting Uehling type potential for a point-like nucleus is given by
\begin{align}
\begin{split}
    \delta V(r)=&-\frac{2e}{\pi}\sum_{i=1}^7\left[\int_{k_i-1}^{k_i}dq\ j_0(qr)\right.\\
    &\left. \cross[A_i+B_i\ln(1+C_i|q^2|)]\vphantom{\int_{k_i-1}^{k_i}} \right].
\end{split}
\end{align}
As stated in Ref.~\cite{Eugen}, using only the parameters of the first momentum region up to infinity is accurate enough at least up to $Z=96$. The Uehling potential then simplifies to
\begin{align}
    \begin{split}
        \delta V_{\text{point}}^{\text{approx}}(r)&=-\frac{2Z\alpha}{\pi}\int_{0}^{\infty}dq\ j_0(qr)[B_1\ln(1+C_1|q^2|)] \\
        &=\frac{2Z\alpha}{\pi}B_1E_1\left(\frac{r}{\sqrt{C_1}}\right),
    \end{split}
    \label{eq:PLApprox}
\end{align}
with the exponential integral
\begin{align}
    E_n(x)=\int_1^{\infty}dt\frac{e^{-xt}}{t^n}.
\end{align}
Inserting this potential into the expression for extended nuclei in Eq.~(\ref{eq:ExtUeh}) results in~\cite{Eugen}
\begin{align}
        \delta V_{\text{fns}}(r)&=-\frac{4\pi e B_1\sqrt{C_1}}{r}\int_{0}^{\infty}dx\ x\rho(x)D_2^{-}(r,x)
\end{align}
with
\begin{align}
    D_n^{\pm}(r,x)=E_n\left(\frac{|r-x|}{\sqrt{C_1}}\right)\pm\left(\frac{|r+x|}{\sqrt{C_1}}\right).
\end{align}
Depending on the nuclear model, we can insert Eq.~(\ref{eq:sphere}) or Eq.~(\ref{eq:fermi}) into $\rho(x)$.

\section{Hyperfine Structure}
Hyperfine splitting occurs due to the magnetic moment generated by the spin of the nucleus $\vb*{\mu_I}$ interacting with the magnetic field $\vb*{B_J}$ caused by the cycling electron. Using the magnetic potential of a point-like magnetic dipole
\begin{align}
    \vb*{A}(\vb*{r})=\frac{\vb*{\mu_I}\cross \vb*{r}}{4\pi r^3}
\end{align}
in order to describe the interaction Hamiltonian, we can arrive at the energy shift resulting from this interaction for an electron bound in a state $|\psi\rangle$ as stated in Ref.~\cite{Beier}
\begin{align}
\begin{split}
\Delta E=&\langle n\kappa m | V_{\text{mag}} | n\kappa m\rangle\\
    =&\langle n\kappa m | e\ \vb*{\alpha}\cdot\vb*{A}(\vb*{r}) | n\kappa m\rangle\\
    =&\frac{e}{4\pi}g_I\mu_N\frac{4\kappa}{4\kappa^2-1}\left(F(F+1)-I(I+1)-j(j+1)\right)\\\
    &\cross \int_0^{\infty}f_{n\kappa}(r)g_{n\kappa}(r)dr
\end{split}
\label{eq:HFSShift}
\end{align}
with $F\in \{ |I-j|,...,I+j\}$. This integral can be evaluated for point-like nuclei, resulting in the following expression:
\begin{align}
\begin{split}
    \Delta E=&\alpha g_I\frac{m_e}{m_p}\frac{\left(F(F+1)-I(I+1)-j(j+1)\right)}{2j(j+1)}\\\
    &\cross m_ec^2\frac{(Z\alpha)^3}{n^3(2l+1)}A(Z\alpha)
    \label{eq:nonrelE}
\end{split}
\end{align}
with the relativistic factor further defined in Ref.~\cite{Pyykko:1973}. In the case of the $1S_{1/2}$ and $2S_{1/2}$ state it is given by
\begin{align}
    A_{1S_{1/2}}(Z\alpha)=&\frac{1}{\gamma(2\gamma-1)},\\
    A_{2S_{1/2}}(Z\alpha)=&\frac{2\left(2(1+\gamma)+\sqrt{2(1+\gamma)}\right)}{(1+\gamma)^2\gamma(4\gamma^2-1)},
\end{align}
with $\gamma=\sqrt{\kappa^2-(Z\alpha)^2}$.\\
This does not, however, account for corrections from QED or for nuclear properties resulting from finite nuclear size when calculating the energy shift. For ground-state electrons in hydrogen-like ions, this can be parameterized as
\begin{align}
    \Delta E = \Delta E_D\left(1-\delta\right)\left(1-\varepsilon\right)+\Delta E_{\text{QED}}
\end{align}
where $\delta$ accounts for the finite size of the nuclear charge distribution and $\varepsilon$ for the finite size of the magnetization distribution, also called the Bohr-Weisskopf correction. $\Delta E_{D}$ is the relativistic Dirac value of hyperfine splitting for point-like nuclei and $\Delta E_{\text{QED}}$ encompasses the possible corrections from QED. For hydrogen-like ions, these corrections have already been determined in Ref.~\cite{Beier,Shabaev1994,Volotka2003,Yerokhin2010oneloop}.\\
Inserting the parametrization into Eq.~(\ref{eq:nonrelE}) results in the following representation\footnote{Additionally there would be a factor $\mathcal{M}$ accounting for finite nuclear mass if following the parametrization given in Ref.\cite{Beier}. It will however cancel out at a later time.}
\begin{align}
    \begin{split}
    \Delta E=&\alpha g_I\frac{m_e}{m_p}\frac{F(F+1)-I(I+1)-j(j+1)}{2j(j+1)} m_ec^2\\
    &\cross \frac{(Z\alpha)^3}{n^3(2l+1)}\left(A(Z\alpha)(1-\delta)(1-\varepsilon)+\frac{\alpha}{\pi}\mathcal{\varepsilon}_{\text{QED}}\right)
    \end{split}
    \label{eq:HFSconv}
\end{align}
where the convention $\varepsilon_{\text{QED}}$ is used to parameterize QED corrections. It can be determined by comparing Eq.~(\ref{eq:HFSShift}) and Eq.~(\ref{eq:HFSconv}).

\section{Calculation Approaches}
\begin{figure}[h!]
    \centering
    \begin{subfloat}
        \centering
        \includegraphics[height=4cm]{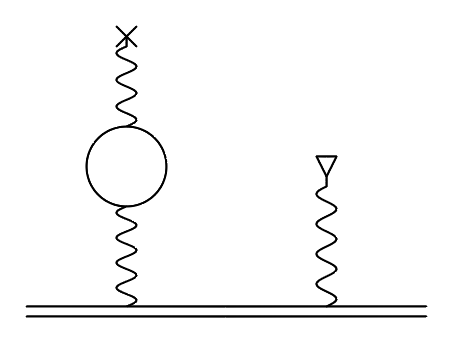}
    \end{subfloat}
    \begin{subfloat}
        \centering
        \includegraphics[height=4cm]{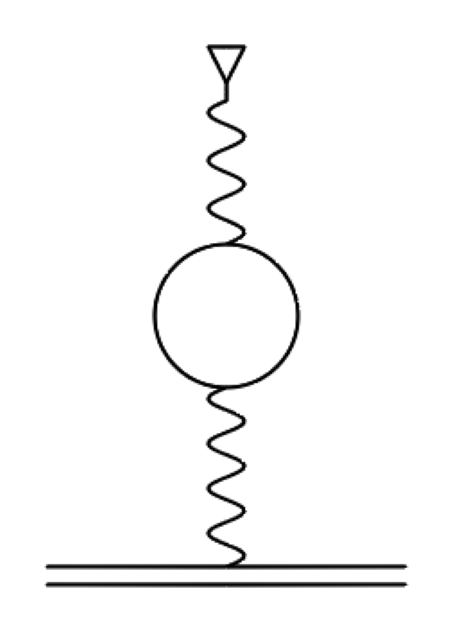}
    \end{subfloat}
    \caption{Vacuum polarization correction to the wave function (left) and the potential (right) to the energy level of a bound electron (represented by the double line) in case of hyperfine splitting. The cross indicates the polarization loop interacting with the Coulomb potential of the nucleus, and the triangle the interaction with the magnetic field of the nucleus. In this case the circle represents both leptonic and hadronic vacuum polarization%\protect\footnotemark
    }
    \label{fig:2ndOrdDiag}
\end{figure}

There are two different ways hyperfine splitting and vacuum polarization can interact at second order which can be seen in Fig.~\ref{fig:2ndOrdDiag}, where vacuum polarization either impacts the wave function or changes the magnetic potential directly.
\renewcommand{\thempfootnote}{\arabic{mpfootnote}}
%\footnotetext[2]{In this case the circle represents both leptonic and hadronic vacuum polarization}

\subsection{Correction to the Wave Function}
In the first case the wave function directly interacts with both the Uehling potential and the magnetic potential. At the second order of perturbation theory, we can, therefore, write the energy shift as
\begin{align}
    \Delta E_{ns}^{(2)} &= 2 \sum_{n' \atop n' \neq n}\frac{\langle ns|\delta V_{\text{Ueh}}|n's\rangle\langle n's|\delta V_{\text{mag}}|ns\rangle}{E_{ns}-E_{n's}}
\end{align}
where we sum over all possible principle quantum numbers of the electron's wave function in between it's interaction with the Uehling potential and the magnetic potential. The factor of 2 arises due to an identical contribution with the potentials interchanged. We can write this expression as
\begin{align}
    \Delta E_{ns}^{(2)} := 2 \langle\psi_{ns}|\delta V_{\text{Ueh}}|\delta\psi_{\text{mag}}\rangle
    \label{eq:plShabaev}
\end{align}
where we have pulled the sum into the expression $\delta\psi_{\text{mag}}$. It describes a magnetically disturbed wave function and for point-like nuclei, following the derivation in Ref.~\cite{ShabaevVirial}, it can be written as
\begin{align}
    |\delta\psi_{\text{mag}}\rangle=\begin{pmatrix}
        iX_{\text{mag}}(r)\Omega_{\kappa m}\\
        -Y_{\text{mag}}(r)\Omega_{\kappa m}
    \end{pmatrix}.
\end{align}
with the radial wave functions
\begin{align}
\begin{split}
     X_{mag}=&\frac{1}{a^2+b}\left(\frac{a}{r}F(r)+\frac{2a\kappa}{r}F(r)+\frac{b}{r}G(r)\right. \\
     &\left. -\frac{b}{\kappa}F(r)(E_{Som}+m)-\frac{a^3\kappa m}{4N^3\gamma}G(r)\vphantom{\frac{a}{r}}\right) \\
     & -\frac{a\left(2E_{Som}-\frac{m}{\kappa}\right)}{a^2+b}\Tilde{G}(r)
\end{split}\\
\begin{split}
     Y_{mag}=&\frac{1}{a^2+b}\left(\frac{a}{r}G(r)-\frac{2a\kappa}{r}G(r)-\frac{b}{r}F(r)\right. \\
     &\left. -\frac{b}{\kappa}G(r)(-E_{Som}+m)-\frac{a^3\kappa m}{4N^3\gamma}F(r)\vphantom{\frac{a}{r}}\right)\\
     &-\frac{a\left(2E_{Som}-\frac{m}{\kappa}\right)}{a^2+b}\Tilde{F}(r)
\end{split}\\
\end{align}
and the derivatives
\begin{align}
    \frac{\partial}{\partial\kappa}|n\kappa\rangle=\begin{pmatrix} \Tilde{G}_{ns}(r)\\ \Tilde{F}_{ns}(r) \end{pmatrix},
\end{align}
which where deterined for 1s and 2s electrons in Ref.~\cite{ShabaevVirial}.\\
Using the convention defined in Eq.~(\ref{eq:HFSconv}), the correction to the energy level can be described with
\begin{align}
    \varepsilon^{\text{wf,pl}}=-\frac{\pi}{\alpha}\frac{\int dr V_{\text{Ueh}}(r)\left(G_{n\kappa}(r)X_{\text{mag}}(r)+F_{n\kappa}(r)Y_{\text{mag}}(r)\right)}{m_e^2(Z\alpha)^3}.
    \label{eq:Enshift}
\end{align}
For extended nuclei, we calculate the correction numerically using B-Splines~\cite{Bspline1, Bspline2}. The approach is similar, however the perturbed wave function is computed with the Uehling potential as the perturbation potential
\begin{align}
    \Delta E_{1s}^{(2)} := 2 \langle\psi_{1s}|V_{\text{mag}}|\delta\psi_{\text{Ueh}}\rangle
\end{align}
with the magnetic potential $V_{\text{mag}}$ defined in~\ref{eq:HFSShift}
The correction to the energy level can then be determined analogously to Eq.~(\ref{eq:Enshift}) with
\begin{align}
    \varepsilon^{\text{wf,fns}}=-\frac{\pi}{\alpha}\frac{\int dr V_{\text{mag}}(r)\left(G_{n\kappa}(r)X_{\text{Ueh}}(r)+F_{n\kappa}(r)Y_{\text{Ueh}}(r)\right)}{m_e^2(Z\alpha)^3}.
\end{align}
Furthermore, we considered Bohr-Weisskopf corrections to the magnetic potential to achieve a more realistic model. We did so by including an additional factor $(r/r_0)^3$, which corrects the point-like magnetic distribution of the nucleus to that of a homogeneous sphere. The results however, did not differ from each other within our margin of error.

\subsection{Correction to the Potential}
When correcting the potential directly we can introduce a factor $F_{\text{ML}}(r)$ to account for the magnetic loop giving us the potential
\begin{align}
    \vb*{A}_{\text{ML}}(\vb*{r})=\frac{\vb*{\mu_I}\cross \vb*{r}}{4\pi r^3} F_{\text{ML}}(r)
\end{align}
For leptonic vacuum polarization, this factor is given in Ref.~\cite{magLoopLep} by
\begin{align}
\begin{split}
    F_{\text{ML,lep}}(r)=&\frac{2\alpha}{3\pi}\int_1^{\infty}dz\sqrt{1-\frac{1}{z^2}}\left(1+\frac{1}{2z^2}\right)\\
    &\cross\frac{1}{z}e^{-2mrz}(2mrz+1).
    \label{eq:ALep}
\end{split}
\end{align}
which was determined by writing the modified potential as
\begin{align}
    \vb*{A}_{\text{ML}}(\vb*{r})=-i\vb*{\mu_I}\cross\left(i\nabla_{\vb*{s}}\int\frac{d^3q}{(2\pi)^3}e^{i\vb*{q}(\vb*{r}-\vb*{s})}\frac{\Pi^R(-q^2)}{q^2}\right)\biggr\rvert_{\vb*{s}=0}.
    \label{eq:corrpotAllg}
\end{align}
In case of hadronic vacuum polarization, we use a similar approach and can carry out the angular integration in Eq.~(\ref{eq:corrpotAllg}). With $R=|\vb*{R}|=|\vb*{r}-\vb*{s}|$ and $q=|\vb*{q}|$ and following Eq.~(\ref{eq:PLApprox}), we can write Eq.~(\ref{eq:corrpotAllg}) as
\begin{align}
    \vb*{A}_{\text{ML,had}}(\vb*{r})=-i\vb*{\mu_I}\cross\left(i\nabla_{\vb*{s}}\frac{B_1}{2\pi R}\int_1^{\infty}dt \frac{e^{\frac{R}{\sqrt{C_1}}t}}{t}\right)\biggr\rvert_{\vb*{s}=0}.
\end{align}
This results in the factor
\begin{align}
    F_{\text{ML,had}}(r)=2B_1\left(e^{-\frac{r}{\sqrt{C}}}+E_1\left(\frac{r}{\sqrt{C_1}}\right)\right)
    \label{eq:AHad}
\end{align}
where we renamed $R\rightarrow r$.
Since both factors are spherically symmetric the angular integration stays the same and the energy correction is given by
\begin{align}
    \varepsilon^{\text{pot}}=\frac{\pi}{\alpha}\frac{\int dr\frac{1}{r^2} G_{n\kappa}(r)F_{n\kappa}(r)F_{\text{ML}}(r)}{m_e^2(Z\alpha)^3}.
    \label{eq:enshiftpot}
\end{align}
This correction was calculated for the bound electron's wave functions of point-like, homogeneous sphere and Fermi distributed nuclei. Again we considered how the results change under consideration of Bohr-Weisskopf effects, resulting in
\begin{align}
    \varepsilon^{\text{pot}}=\frac{\pi}{\alpha}\frac{\int dr\frac{r}{r_0} G_{n\kappa}(r)F_{n\kappa}(r)F_{\text{ML}}(r)}{m_e^2(Z\alpha)^3}.
    \label{eq:enshiftpot}
\end{align}
The results, again, did not differ within our margin of error.\\

Using the nuclear radii listed in Tab.~\ref{tab:rad}, we can calculate the values of $\varepsilon$ for corrections to the wave function (wf) and the magnetic potential (mag) in case of point-like nuclei (pl) and extended nuclei (fns) for the discussed nuclear models (sphere, Fermi). The uncertainty of the given values arises from the uncertainty of the rms radius of the isotopes. The results in case of 1s and 2s states can be seen in Tab.~[\ref{tab:ElVP1s}-\ref{tab:HadVP2s}]. When calculating the 2s states, we have to account for a factor of 8 due to Eq.~(\ref{eq:HFSconv}) being proportional to $1/n^3$. Furthermore, literature values by Beier~\cite{Beier} and Karshenboim et al.~\cite{Karshenboim} are given for comparisons, when available. For fns results, Beier sometimes used a spherical model and sometimes a Fermi model. We have consistently calculated both models from $Z\geq 4$. In section \ref{sec:summ} we discuss the results in more detail.\\
\begin{table}[h!]
    \centering
    \setlength\tabcolsep{0pt}
    \caption{Studied isotopes, their rms radii and their uncertainty \cite{AtomicData}}
    \begin{tabular*}{\linewidth}{@{\extracolsep{\fill}} cccc }
    \hline
    \hline
        Nuclear Charge $Z$ &  Mass Number $A$ & $R_{\text{rms}}$(fm) & $\Delta R$(fm)\\
        \hline
        1 & 1 & 0.8783 & 0.0086\\
        2 & 4 & 1.6755 & 0.0028\\
        4 & 9 & 2.5190 & 0.0120\\
        5 & 10 & 2.4277 & 0.0499\\
        10 & 20 & 3.0055 & 0.0021\\
        25 & 55 & 3.7057 & 0.0022\\
        44 & 104 & 4.5098 & 0.0020\\
        63 & 145 & 4.9663 & 0.0091\\
        70 & 176 & 5.3215 & 0.0062\\
        83 & 209 & 5.5211 & 0.0026\\
        92 & 238 & 5.8571 & 0.0033\\
        \hline
        \hline
    \end{tabular*}
    \label{tab:rad}
\end{table}\\

\section{Analytic Approximations}
For point-like nuclei and small charge numbers, we can use approximate formulas to compare to our results as done in Ref.~\cite{ApproxFrom}. We do so by first approximating the free and magnetically perturbed radial wave functions, as well as their product, for point-like nuclei in the lowest order of $Z\alpha$, instead of using the full relativistic expressions:
\begin{align}
    \begin{split}
        G(r)&\approx r2(m_eZ\alpha)^{\frac{3}{2}}e^{-m_eZ\alpha r},\\
        F(r)&\approx 0,\\
        X_{\text{mag}}(r)&\approx G(r),\\
        G(r)F(r)&\approx -2m_e^3(Z\alpha)^4e^{-2\lambda r}.
    \end{split}
    \label{eq:ApproxWF}
\end{align}
Inserting these into Eq.~(\ref{eq:Enshift}) and Eq.~(\ref{eq:enshiftpot}) results in the same expression for both approaches. For leptonic VP, it is given by
\begin{align}
    \varepsilon^{\text{lep,pl}}_{\text{approx}}=\frac{3}{8}\pi Z\alpha\frac{m_e}{m_{\text{lep}}},
    \label{eq:approxPL1}
\end{align}
to which our results for electronic VP coincide within the first digit up until $Z=5$ for the corrected wave function and $Z=25$ for the corrected magnetic potential. For muonic VP, it coincides up until $Z=25$ in both cases. For hadronic VP, it is given by
\begin{align}
    \varepsilon^{\text{had,pl}}_{\text{approx}}=8\pi m_eZB_1\sqrt{C_1},
    \label{eq:approxPL2}
\end{align}
coinciding within the first digit up until $Z=4$. Given the different approach used to obtain these approximations, the similarity to our exact numerical results indicates their legitimacy.\\
To find an analytic approximation for extended nuclei, we inserted the Uehling potential for a spherical nucleus and wave functions for a point-like nucleus in our numerical calculations. For small nuclear charge numbers the results are very similar to the numerical fns results in case of leptonic VP. This allowed us to derive an analytic approximation formula for extended nuclei, by using the wave functions in Eq.~(\ref{eq:ApproxWF}) and the expression for an extended Uehling potential given in Eq.~(\ref{eq:ExtUeh}). Still considering small charge numbers, the results also coincide within the first digit, as can be seen in Tab.~[\ref{tab:ElVP1s},\ref{tab:MuVP1s}]. This approach does not work in case of hadronic vacuum polarization, but, as expected, the results for hadronic VP behave similarly to those of muonic VP in regards of magnitude.\\
Furthermore, we can compare our results to the approximation formula for small nuclei given in Ref.~\cite{KarshenboimApprox}:
\begin{align}
    \varepsilon^{\text{wf,fns}}_{\text{el}}= \frac{70}{12\sqrt{12}}Z\alpha m_e r_{\text{rms}}\left(2\ln{r_{\text{rms}}m_e-\ln{12}+\frac{1902}{630}}\right).
\end{align}
The results for electronic VP lie within the same magnitude. Considering the difference between electronic and muonic VP results in the following expression:
\begin{align}
    \varepsilon^{\text{wf,fns}}_{\text{mu}}-\varepsilon^{\text{wf,fns}}_{\text{el}}=-C Z\alpha m_e r_\text{rms}\ln{\frac{m_\mu}{m_e}}.
\end{align}
with a constant $C=0.0003393$ defined by this difference. When considering the difference of our two numerical results, we can see that for smaller nuclei the value for $C$ stays constant within the first digit at $C=0.0002$. This deviates slightly from the theoretical value, however, due to the very rough approximation the important aspect is that it stays constant for the various nuclear sizes.

\section{Summary}
\label{sec:summ}
In this work, corrections to hyperfine splitting induced by the leptonic and hadronic Uehling potential were computed by correcting the wave function of the bound electron and by correcting the nuclear potential directly. They were determined for hydrogen-like systems from $Z=1$ to $Z=92$ for point-like, spherical and Fermi distributed nuclei. For extended nuclei a B-Spline representation of Dirac wave functions employing dual kinetic balance was used to compute the results. In case of electronic VP, all results coincided well with the literature values available. Analytic approximations for point-like nuclei and in case of leptonic VP for fns nuclei were determined and agreed with our exact numerical results. They were further reassured by using an approximation derived previously by Karshenboim. Our results also support the claim in this paper, that for muonic VP one can use 30\% of the point-like values to determine the fns values.\\
In Table~\ref{tab:Z4} corrections for $Z=4$ in case of $1s$ nuclei are shown. The complete results can be found in the appendix.
\begin{table}[h!]
    \centering
    \setlength\tabcolsep{0pt}
    \caption{Vacuum Polarization in Beryllium}
    \begin{tabular*}{\linewidth}{@{\extracolsep{\fill}} cccc }
    \hline
    \hline
        & Point-Like &  Sphere & Fermi\\
        \hline
        $\varepsilon_{\text{el}}^{\text{wf}}$ & 0.0366463 & 0.0353555(50) & 0.0354155(54)\\
        $\varepsilon_{\text{el}}^{\text{pot}}$ & 0.0338856 & 0.0326152(49) & 0.0326751(53)\\
        \hline
        $\varepsilon_{\text{mu}}^{\text{wf}}$ & 0.00016827  & 0.00003060(13) & 0.00003413(17) \\
        $\varepsilon_{\text{mu}}^{\text{pot}}$  & 0.00016786  & 0.00003049(13) & 0.00003402(18)\\
        \hline
        $\varepsilon_{\text{had}}^{\text{wf}}$  & 0.000239897 & 0.000021458(99) & 0.00002462(15)\\
        $\varepsilon_{\text{had}}^{\text{pot}}$ & 0.000239402 & 0.000021386(99) & 0.00002455(15)\\
        \hline
        \hline
    \end{tabular*}
    \label{tab:Z4}
\end{table}\\
One can see that the results for muonic and hadronic VP are strongly suppressed compared to electronic vacuum polarization as expected, since in case of leptonic VP the Uehling potential decreases with increasing mass of the loop particle as seen in eq.~\ref{eq:approxPL1}. Furthermore, hadronic VP is often said to have the same magnitude as muonic VP, since the mass of the lightest hadron, namely the pion, is similar to that of a muon.\\
Further, we can see that the results are extremely sensitive to the nuclear model used. In fact, the model used leads to a greater discrepancy in the results than the uncertainty in the rms radius, even when comparing the fns models to each other. This discrepancy between fns models can also be seen in $2s$ nuclei. There is a slight trend, that it is greater when considering electronic VP relatively compared to $1s$ nuclei but smaller for muonic and hadronic VP. However, in both cases it lies within the uncertainty of the results and could thus be meaningless. It was expected that the solutions for point-like nuclei would differ more with increasing $Z$. However, for muonic and hadronic vacuum polarization there is already a discrepancy in the first digit between point-like and fns solutions and for higher Z the results differ by multiple magnitudes. This suggests that the vacuum polarization itself might already take place within the nucleus, or at least be significantly closer to it, than when considering corrections due to a homogeneous external field.\\

These results, therefore, show that when considering hyperfine effects, it is necessary to have a better understanding of the nuclear model.\\
Experiments to better understand the nuclear structure by using muonic atoms are planned at the Paul Scherrer Institut. Furthermore, the methods introduced in Ref.~\cite{Igor} could reduce nuclear uncertainties due to charge distribution dependence, making our results relevant when considering experimental data. It would also be interesting to consider hadronic vacuum polarization through a QCD approach.

%%%%%%%%%%%%%%%%%%%%%%%%%%%%%%%%%%%%%%%%%%%%%%%%%%%%%%%%%%%%%%%%%%%%%%%%%%%

\onecolumngrid
\section*{Acknowledgments}
\noindent We thank N. Oreshkina and Z. Harman for the insightful discussions and  H. Cakir and S. Banerjee for providing the B-Spline code used and modified in the calculations.\\
This work is supported by the Deutsche Forschungsgemeinschaft (DFG, German Research Foundation) Project-ID 273811115- SFB 1225. It comprises parts of the BSc thesis work of J.H., submitted to Heidelberg University, Germany.

\twocolumngrid

\onecolumngrid
\appendix*
\newpage
\section{List of Tables}
\label{appendix}
\begin{table*}[h!]
    \setlength\tabcolsep{0pt}
    \caption{Electronic vacuum polarization corrections for $1s$ states.}
    
    \begin{tabular*}{\linewidth}{@{\extracolsep{\fill}} llllllll }
    \hline
    \hline
        Z & $\varepsilon_{\text{el}}^{\text{wf,pl}}$ & $\varepsilon_{\text{el}}^{\text{pot,pl}}$ & $\varepsilon_{\text{el,sphere}}^{\text{wf,fns}}$ & $\varepsilon_{\text{el},\text{fermi}}^{\text{wf,fns}}$ & $\varepsilon_{\text{el},\text{sphere}}^{\text{pot,fns}}$ & $\varepsilon_{\text{el},\text{fermi}}^{\text{pot,fns}}$ & Refs.\\
        \hline
        1 & 0.0087703 & 0.0085579 & 0.0086376(11) & & 0.0084255(12)\\
         & 0.0087691 & 0.0085578 & & & & & \cite{Beier} \\
         & 0.0087703 & 0.0085578 & & & & & \cite{Karshenboim} \\
         & 0.0085970 & 0.0085970 & 0.00846474(11) & & & & approx \\
        2 & 0.01781815 & 0.01704863 & 0.01736232(63) & & 0.01659564(62)\\
         & 0.017817 & 0.017049 & & & & & \cite{Beier} \\
         & 0.01719398 & 0.01719398 & 0.01674208(62) & & & & approx \\
        4 & 0.0366463 & 0.0338856 & 0.0353555(50) & 0.0354155(54) & 0.0326152(49) & 0.0326751(53)\\
         & 0.036645 & 0.033886 & & & & & \cite{Beier} \\
         & 0.0343880 & 0.0343880 & 0.0331290(48) & & & & approx \\
        5 & 0.046415 & 0.042257 & 0.044833(26) & 0.044913(29) & 0.040708(26) & 0.040787(28)\\
         & 0.046414 & 0.042257 & & & & & \cite{Beier} \\
         & 0.0464153 & 0.0422566 & & & & & \cite{Karshenboim} \\
         & 0.042985 & 0.042985 & 0.041457(25) & & & & approx \\
        10 & 0.0989572 & 0.0840724 & 0.0949078(23) & 0.0950343(24) & 0.0802593(21) & 0.0803801(22)\\
         & 0.098955 & 0.084072 & 0.094922 & & 0.080275 & & \cite{Beier} \\
         & 0.0989572 & 0.0840724 & & & & & \cite{Karshenboim} \\
         & 0.0859699 & 0.0859699 & 0.0823433(20) & & & & approx \\
        25 & 0.3068919 & 0.2192739 & 0.2887396(81) & 0.2890727(84) & 0.2051510(64) & 0.2054255(66)\\
         & 0.3068919 & 0.2192738 & & & & & \cite{Karshenboim} \\
         & 0.2149248 & 0.2149248 & 0.2042159(50) & & & & approx \\
        44 & 0.794009 & 0.458597 & 0.706157(25) & 0.707029(26) & 0.409936(15) & 0.410497(15)\\
         & & & 0.7065 & & 0.4102 & & \cite{Beier} \\
        63 & 2.04367 & 0.91404 & 1.59915(42) & 1.60183(42) & 0.74805(18) & 0.74932(18)\\
         & & & & 1.59 & & 0.73 & \cite{Beier} \\
        70 & 3.04459 & 1.21160 & 2.17126(49) & 2.17517(47) & 0.93193(17) & 0.93360(18)\\
        83 & 7.43667 & 2.24726 & 4.03804(52) & 4.04772(53) & 1.45440(15) & 1.45768(15)\\
         & & & 4.038 & 4.05 & 1.455 & 1.46 & \cite{Beier} \\
        92 & 16.9325 & 3.92182 & 6.3778(13) & 6.3956(14) & 2.01655(33) & 2.02169(33)\\
         & & & 6.377 & & 2.016 & & \cite{Beier} \\
        \hline
        \hline
        \label{tab:ElVP1s}
    \end{tabular*}

    \centering
    \caption{Muonic vacuum polarization corrections for $1s$ states.}
    \begin{tabular*}{\linewidth}{@{\extracolsep{\fill}} llllllll }
    \hline
    \hline
        Z & $\varepsilon_{\text{mu}}^{\text{wf,pl}}$ & $\varepsilon_{\text{mu}}^{\text{pot,pl}}$ & $\varepsilon_{\text{mu},\text{sphere}}^{\text{wf,fns}}$ & $\varepsilon_{\text{mu},\text{fermi}}^{\text{wf,fns}}$ & $\varepsilon_{\text{mu},\text{sphere}}^{\text{pot,fns}}$ & $\varepsilon_{\text{mu},\text{fermi}}^{\text{pot,fns}}$ & Refs.\\
        \hline
        1 & 0.00004162 & 0.000041604 & 0.00001655(10) & & 0.000016543(98)\\
        & 0.000041578 & 0.000041578 & 0.000016535(96) & & & & approx \\
        2 & 0.000083427 & 0.000083359 & 0.000021298(28) & & 0.000021269(28)\\
        & 0.000083156 & 0.000083156 & 0.000021229(28) & & & & approx \\
        4 & 0.00016827 & 0.00016786 & 0.00003060(13) & 0.00003413(17) & 0.00003049(13) & 0.00003402(18)\\
        & 0.00016631 & 0.00016631 & 0.00003028(13) & & & & approx \\
        5 & 0.00021161 & 0.00021086 & 0.00003968(74) & 0.0000446(10) & 0.00003951(72) & 0.0000444(10)\\
        & 0.00020789 & 0.00020789 & 0.00003908(70) & & & & approx \\
        10 & 0.000444046 & 0.000438598 & 0.000067984(43) & 0.000073562(54) & 0.000067246(43) & 0.000072809(54)\\
        & 0.000415779 & 0.000415779 & 0.000064687(41) & & & & approx \\
        25 & 0.00151814 & 0.001413614 & 0.00017115(10) & 0.00018059(12) & 0.000164787(95) & 0.00017410(11)\\
        & 0.00103945 & 0.00103945 & 0.000133332(74) & & & & approx \\
        44 & 0.00573678 & 0.00458894 & 0.00038522(18) & 0.00039987(20) & 0.00035179(17) & 0.00036593(19)\\
        63 & 0.0283835 & 0.0174905 & 0.0009377(20) & 0.0009676(22) & 0.0007954(17) & 0.0008233(19)\\
        70 & 0.0579742 & 0.0312626 & 0.0012740(18) & 0.0013098(20) & 0.0010441(15) & 0.0010769(17)\\
        83 & 0.2880179 & 0.1119933 & 0.0026262(17) & 0.0026961(18) & 0.0020019(13) & 0.0020633(14)\\
        92 & 1.1990217 & 0.3390947 & 0.0043050(36) & 0.0044087(38) & 0.0030906(26) & 0.0031785(28)\\
        \hline
        \hline
        \label{tab:MuVP1s}
    \end{tabular*}
\end{table*}

\begin{table*}[h]
    \caption{Hadronic vacuum polarization corrections for $1s$ states.}
    \begin{tabular*}{\linewidth}{@{\extracolsep{\fill}} llllllll }
    \hline
    \hline
        Z & $\varepsilon_{\text{had}}^{\text{wf,pl}}$ & $\varepsilon_{\text{had}}^{\text{pot,pl}}$ & $\varepsilon_{\text{had},\text{sphere}}^{\text{wf,fns}}$ & $\varepsilon_{\text{had},\text{fermi}}^{\text{wf,fns}}$ & $\varepsilon_{\text{had},\text{sphere}}^{\text{pot,fns}}$ & $\varepsilon_{\text{had},\text{fermi}}^{\text{pot,fns}}$ & Refs.\\
        \hline
        1 & 0.00005931 & 0.00005930 & 0.00001385(11) & & 0.00001384(11)\\
        & 0.00005926 & 0.00005926 & & & & & approx \\
        2 & 0.000118901 & 0.000118827 & 0.000015720(25) & & 0.000015700(25)\\
        & 0.000118516 & 0.000118516 & & & & & approx \\
        4 & 0.000239897 & 0.000239402 & 0.000021458(99) & 0.00002462(15) & 0.000021386(99) & 0.00002455(15)\\
        & 0.000237031 & 0.000237031 & & & & & approx \\
        5 & 0.00030178 & 0.00030085 & 0.00002793(56) & 0.00003242(85) & 0.00002781(57) & 0.00003230(90)\\
        & 0.00029629 & 0.00029629 & & & & & approx \\
        10 & 0.000634961 & 0.000627692 & 0.000046964(32) & 0.000051587(43) & 0.000046489(32) & 0.000051104(42)\\
        & 0.000592579 & 0.000592579 & & & & & approx \\
        25 & 0.002215687 & 0.002066843 & 0.000116714(70) & 0.000123978(84) & 0.000112649(69) & 0.000119854(82)\\
        & 0.001481446 & 0.001481446 & & & & & approx \\
        44 & 0.00883423 & 0.00708513 & 0.00026054(13) & 0.00027125(14) & 0.00023910(12) & 0.00024956(13)\\
        63 & 0.0477348 & 0.0294996 & 0.0006323(14) & 0.0006538(15) & 0.0005407(12) & 0.0005611(13)\\
        70 & 0.1017223 & 0.0550071 & 0.0008576(13) & 0.0008830(13) & 0.0007090(11) & 0.0007327(12)\\
        83 & 0.5559891 & 0.21670903 & 0.0017666(12) & 0.0018158(13) & 0.00136180(92) & 0.0014062(10)\\
        92 & 2.5102986 & 0.7112905 & 0.0028925(25) & 0.0029648(27) & 0.0021021(19) & 0.0021652(20)\\
        \hline
        \hline
        \label{tab:HadVP1s}
    \end{tabular*}
    \label{tab:VP1s}

    \centering
    \setlength\tabcolsep{0pt}
    \caption{Electronic vacuum polarisation corrections for $2s$ states.}
    \begin{tabular*}{\linewidth}{@{\extracolsep{\fill}} lllllll }
    \hline
    \hline
        Z & $\varepsilon_{\text{el}}^{\text{wf,pl}}$ & $\varepsilon_{\text{el}}^{\text{pot,pl}}$ & $\varepsilon_{\text{el},\text{sphere}}^{\text{wf,fns}}$ & $\varepsilon_{\text{el},\text{fermi}}^{\text{wf,fns}}$ & $\varepsilon_{\text{el},\text{sphere}}^{\text{pot,fns}}$ & $\varepsilon_{\text{el},\text{fermi}}^{\text{pot,fns}}$ \\
        \hline
        1 & 0.0087480 & 0.0085635 & 0.0086153(11) & &  0.0084312(11) \\
        2 & 0.01773110 & 0.01707253 & 0.01727519(63) & & 0.01661945(62)\\ 
        4 & 0.0363164 & 0.0339916 & 0.0350248(50) & 0.0350848(54) & 0.0327203(49) & 0.0327802(53)\\
        5 & 0.045914 & 0.042430 & 0.044330(26) & 0.044410(29) & 0.040880(26) & 0.040959(28)\\
        10 & 0.0972437 & 0.0849339 & 0.0931777(23) & 0.0933049(24) & 0.0811029(22) & 0.0812243(23)\\
        25 & 0.3024963 & 0.2282650 & 0.2838533(83) & 0.2841949(86) & 0.2137208(66) & 0.2140035(68)\\
        44 & 0.824157 & 0.508624 & 0.728278(28) & 0.729229(28) & 0.455278(17) & 0.455892(17)\\
        63 & 2.35860 & 1.11359 & 1.82183(51) & 1.82506(51) & 0.91259(21) & 0.91413(21)\\
        70 & 3.70526 & 1.54155 & 2.59865(58) & 2.60361(59) & 1.18673(22) & 1.18884(22)\\
        83 & 10.18887 & 3.14582 & 5.39983(73) & 5.41345(75) & 2.03183(21) & 2.03643(22) \\
        92 & 25.5485 & 5.94539 & 9.3306(21) & 9.3579(21) & 3.03565(50) & 3.04347(50)\\
        \hline
        \hline
        \label{tab:ElVP2s}
    \end{tabular*}
    \caption{Muonic vacuum polarisation corrections for $2s$ states.}
    \begin{tabular*}{\linewidth}{@{\extracolsep{\fill}} lllllll }
    \hline
    \hline
        Z & $\varepsilon_{\text{mu}}^{\text{wf,pl}}$ & $\varepsilon_{\text{mu}}^{\text{pot,pl}}$ & $\varepsilon_{\text{mu},\text{sphere}}^{\text{wf,fns}}$ &  $\varepsilon_{\text{mu},\text{fermi}}^{\text{wf,fns}}$ & $\varepsilon_{\text{mu},\text{sphere}}^{\text{pot,fns}}$ & $\varepsilon_{\text{mu},\text{fermi}}^{\text{pot,fns}}$ \\
        \hline
        1 & 0.000041619 & 0.00004161 & 0.000016552(96) & & 0.00001644(10)\protect\footnotemark  \\
        2 & 0.000083440 & 0.000083375 & 0.000021300(28) & & 0.000021273(28)\\ 
        4 & 0.00016839 & 0.00016799 & 0.00003062(13) & 0.00003415(18) & 0.00003051(13) & 0.00003404(18) \\
        5 & 0.00021185 & 0.00021111 & 0.00003972(72) & 0.0000447(11) & 0.00003956(71) & 0.0000445(10) \\
        10 & 0.000446079 & 0.000440673 & 0.000068249(44) & 0.000073853(55) & 0.000067575(43) & 0.000073164(54)\\
        25 & 0.00156307 & 0.001455928 & 0.00017587(10) & 0.00018558(12) & 0.000169803(98) & 0.00017940(12)\\
        44 & 0.00628753 & 0.00503132 & 0.00042069(20) & 0.00043675(22) & 0.00038607(18) & 0.00040158(20)\\
        63 & 0.0343610 & 0.0211769 & 0.0011307(24) & 0.0011669(27) & 0.0009641(21) & 0.0009979(23)\\
        70 & 0.0735303 & 0.0396513 & 0.0016097(23) & 0.0016551(25) & 0.0013255(20) & 0.0013671(21)\\
        83 & 0.4044851 & 0.1572363 & 0.0036796(24) & 0.0037776(26) & 0.0028103(18) & 0.0028964(20)\\
        92 & 1.8299643 & 0.5173064 & 0.0065634(55) & 0.0067213(59) & 0.0047050(40) & 0.0048385(43)\\
        \hline
        \hline
        \label{tab:MuVP2s}
    \end{tabular*}

    \begin{@twocolumnfalse}
    $^1$The uncertainty of this value arises due to numerical instability. For all other values it continues to come from the uncertainty in rms radius.
    %\footnotetext{The uncertainty of this value arises due to numerical instability. For all other values it continues to come from the uncertainty in rms radius.}
    \end{@twocolumnfalse}

    \caption{Hadronic vacuum polarisation corrections for $2s$ states.}
    \begin{tabular*}{\linewidth}{@{\extracolsep{\fill}} lllllll }
    \hline
    \hline
        Z & $\varepsilon_{\text{had}}^{\text{wf,pl}}$ & $\varepsilon_{\text{had}}^{\text{pot,pl}}$ & $\varepsilon_{\text{had},\text{sphere}}^{\text{wf,fns}}$ &  $\varepsilon_{\text{had},\text{fermi}}^{\text{wf,fns}}$ & $\varepsilon_{\text{had},\text{sphere}}^{\text{pot,fns}}$ & $\varepsilon_{\text{had},\text{fermi}}^{\text{pot,fns}}$ \\
        \hline
        1 & 0.00005931 & 0.00005930 & 0.00001385(11) & & 0.00001384(11) \\
        2 & 0.000118921 & 0.000118850 & 0.000015721(25) & & 0.000015704(25)\\ 
        4 & 0.00024007 & 0.00023958 & 0.00002147(10) & 0.00002464(15) & 0.00002140(10) & 0.00002457(15) \\
        5 & 0.00030213 & 0.00030120 & 0.00002795(57) & 0.00003245(90) & 0.00002785(56) & 0.00003234(90) \\
        10 & 0.000637911 & 0.000630653 & 0.000047148(32) & 0.000051793(43) & 0.000046716(32) & 0.000051353(43)\\
        25 & 0.002281580 & 0.002128635 & 0.000119937(72) & 0.000127417(86) & 0.000116077(71) & 0.000123497(85)\\
        44 & 0.00968381 & 0.00776779 & 0.00028454(14) & 0.00029629(15) & 0.00026240(13) & 0.00027386(15)\\
        63 & 0.0577908 & 0.0357159 & 0.0007625(17) & 0.0007885(19) & 0.0006554(15) & 0.0006801(16)\\
        70 & 0.1290163 & 0.0697652 & 0.0010836(16) & 0.0011158(17) & 0.0009001(14) & 0.0009302(15)\\
        83 & 0.7807174 & 0.3042536 & 0.0024752(16) & 0.0025442(18) & 0.0019118(13) & 0.0019740(14)\\
        92 & 3.8305248 & 1.0851223 & 0.0044098(38) & 0.0045200(40) & 0.0032003(28) & 0.0032962(30)\\
        \hline
        \hline
        \label{tab:HadVP2s}
    \end{tabular*}
\end{table*}

\end{document}